\documentstyle[aps,prl]{revtex}
\begin{document}

\input epsf.sty
\twocolumn[\hsize\textwidth\columnwidth\hsize\csname %
@twocolumnfalse\endcsname
 
\draft
\widetext

\title{A Monte Carlo Study of Correlations in Quantum Spin Ladders}

\author{M. Greven, R. J. Birgeneau, and U.--J. Wiese}
\address{Department of Physics, Massachusetts Institute of Technology,
Cambridge, MA 02139}

\date{\today}
\maketitle
  
\begin{abstract}

We study antiferromagnetic spin--1/2 Heisenberg ladders, comprised of $n_c$ 
chains ($2 \leq n_c \leq 6$) with ratio $J_{\bot}/J_{\|}$ of inter-- to 
intra--chain couplings.
From measurements of the correlation function we deduce the correlation length 
$\xi(T)$.
For even $n_c$, the static structure factor exhibits a peak at a temperature 
below the corresponding spin gap.
Results for isotropically coupled ladders ($J_{\bot}/J_{\|}=1$) are compared 
to those for the single chain and the square lattice.
For $J_{\bot}/J_{\|} \leq 0.5$, the correlation function of the two--chain 
ladder is in excellent agreement with analytic results from conformal field 
theory, and $\xi(T)$ exhibits simple scaling behavior.

\end{abstract}


\phantom{.}
]

\narrowtext
 
Low--dimensional quantum Heisenberg antiferromagnets (QHA) exhibit many 
fascinating properties.
In 1931, Bethe demonstrated for the one--dimensional (1D) spin $S=1/2$ 
nearest--neighbor Heisenberg chain that quantum fluctuations prevent the 
existence of an ordered ground state \cite{Bet31}.
Instead, this system exhibits power--law correlations with gapless excitations.
Haldane suggested in 1983 that all non--integer--spin chains are gapless, but 
that integer--spin chains should have a spin gap \cite{Hal83}.
By now, there is much 
evidence for the correctness of this famous conjecture \cite{1D}.
For the two--dimensional (2D) analog of the spin chains, the square--lattice 
nearest--neighbor QHA, it has been established over the past decade that an 
ordered ground state exists even in the extreme quantum limit of $S=1/2$ 
\cite{SRCUOCL}.

Spin ladders are arrays of coupled chains, and thus present interpolating 
structures between 1D and 2D \cite{Dag96}.
These systems are thought to be realized, with $S = 1/2$, in the materials 
$(VO)_2 P_2 O_7$ \cite{Joh87} and $Sr_{n-1} Cu_{n+1} O_{2n}$ \cite{Hir91}. 
They allow the study of the dimensional crossover from the power--law 
correlations of the $S=1/2$ chain to the long--range order of the square 
lattice.
Interestingly, ladders with an odd number $n_c$ of chains have power--law spin 
correlations in their ground state, while those with even $n_c$ exhibit 
exponentially decaying correlations due to the presence of a spin gap   
\cite{Dag96}.
In analogy to the general $S$ chain \cite{Hal83}, the fundamental difference 
between even and odd ladders is thought to be of topological nature 
\cite{Whi94,Sie96}.

In this Letter we investigate isotropically coupled ladders as well as the 
two--chain ladder in the regime of weak inter--chain couplings.
The former systems are directly relevant to the known experimental systems 
which are approximately isotropic \cite{Dag96}.
Moreover, the crossover from 1D to 2D is most naturally studied in this case.
We perform numerical simulations which allow us to determine both the 
correlation length and the static structure factor down to very low 
temperatures.
Our results provide a basis for comparison with future neutron scattering and 
NMR experiments.
The two--chain ladder at weak inter--chain couplings allows us to investigate 
the formation of a spin gap away from the unstable $T=0$ fixed point of the 
gapless $S=1/2$ chain \cite{Bar93}, and indeed serves to demonstrate the 
extreme fragility of the $S=1/2$ chain power--law correlations.
Our data for the intra-- and inter--chain correlation functions at low 
temperature are in excellent agreement with a recent theoretical prediction 
\cite{She96}.
Moreover, we discover that in this regime the correlation length exhibits 
universal scaling behavior with respect to the inter--chain coupling.

The Hamilton operator for a Heisenberg system of $n_c$ 
$S = 1/2$ chains forming a ladder is 
\begin{equation}
H = J_{\|} \sum_{\langle ij\rangle_{\|}} {\bf S}_i \cdot {\bf S}_j
  + J_{\bot} \sum_{\langle ij\rangle_{\bot}} {\bf S}_i \cdot {\bf S}_j.
\label{H}
\end{equation}
Here, ${\bf S}_i = \frac{1}{2}{\bf\sigma}_i$ is the quantum spin operator
located at the point $i$, while $\langle ij \rangle_{\|}$ and 
$\langle ij \rangle_{\bot}$ denote nearest neighbors along and between chains, 
respectively. We consider antiferromagnetic couplings, that is $J_{\|}, 
J_{\bot} > 0$, and periodic boundary conditions along the chains.
We use units in which $\hbar = k_B = 1$ and, unless noted otherwise, 
$J_{\|} = 1$.

The ladder systems are investigated with a very efficient loop cluster 
algorithm \cite{Eve93,Wie94} which allows access to very low temperatures and
the implementation of improved estimators in order to reduce statistical errors
of observables. 
We simulate lattices large enough, both along the chains and in Euclidean time,
so that finite size and finite Trotter number effects are comparable to the 
statistical errors.
In particular, the length of the ladders is kept $\sim 15$ times larger than 
the correlation length.
Typically, $4\times10^4$ loop updates are performed for equilibration, 
followed by $4\times10^5$ measurements.

In order to obtain information about the gap $\Delta(n_c,J_{\bot})$ for even 
$n_c$, we measure the uniform susceptibility
\begin{equation}
\chi(n_c,J_{\bot};T) \sim T^{-1} \langle (\sum_i  S_{i} ^{z})^2 \rangle.
\end{equation}
Gap values extracted from fits to the low--$T$ form \cite{Tro94}
\begin{equation}
\chi(n_c,J_{\bot};T) \sim T^{-1/2} e^{-\Delta(n_c,J_{\bot})/T}
\end{equation}
are shown in Fig. 1.
For $J_{\bot} \geq 2$, we find very good agreement with strong--coupling 
expansion results for $n_c =2$ and 4 \cite{Rei94}, shown as solid lines.
Our susceptibility data for the isotropically coupled ladders 
($J_{\bot} = 1$) agree well with recent Monte Carlo work \cite{Fri96}, and
we obtain $\Delta(n_c,1) = 0.502(5), 0.160(5)$, and $0.055(6)$ for 
$n_c = 2,4,$ and $6$, respectively.
In an earlier study \cite{Whi94}, $\Delta(2,1)=0.504$ and 
$\Delta(4,1)=0.190$ were found.
For $n_c = 2$, we are able to access the weak--coupling regime characterized 
by $\Delta(2,J_{\bot}) \sim J_{\bot}$ \cite{She96,Hat94}.
We find $\Delta (2,J_{\bot})/J_{\bot} = 0.41(1)$,
which is somewhat smaller than $\Delta(2,J_{\bot})/J_{\bot} = 0.47(1)$ 
obtained previously \cite{Hat94}.
\begin{figure}
\centerline{\epsfxsize=3.25in\epsfbox
{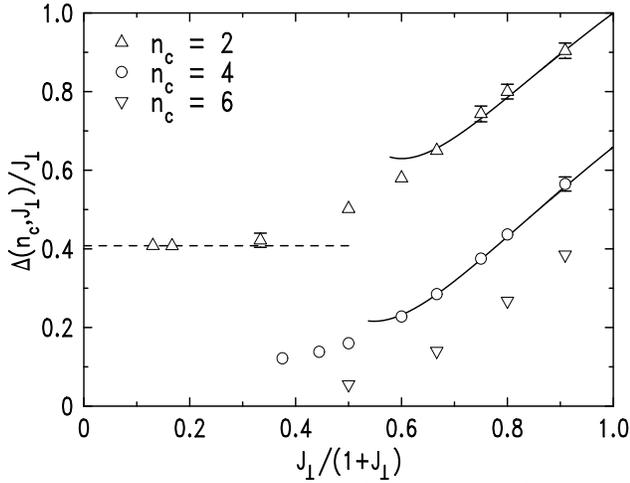}}
\caption{
Dependence of the spin gap on the inter--chain coupling.
The solid lines represent strong--coupling results \protect\cite{Rei94}, and 
the dashed line indicates the weak--coupling behavior 
$\Delta(2,J_{\bot} \rightarrow 0) = 0.41(1)J_{\bot}$.
}
\end{figure}

Next we compute the staggered correlation function 
\begin{equation}
C(i,j) = (-1)^{\mbox{sign}(i,j)} \langle {\bf S}_i \cdot {\bf S}_j \rangle,
\end{equation}
where $\mbox{sign}(i,j)$ is $1 (-1)$ if the spins at $i$ and $j$ are separated 
by an even (odd) number of couplings.
For a system of two weakly coupled
chains at $T = 0$, conformal field theory predicts \cite{She96}
\begin{equation}
C(r) = G_+(r) G_-(r) [G_-(r) G_-(3 r) \pm G_+(r) G_+(3 r)],
\end{equation}
and this form is claimed to be exact in the continuum limit at small $J_{\bot}$.
The sign of the last term is plus for intra-- and minus for inter--chain 
correlations, and $r$ measures the distance of two spins along the ladder.
The functions 
\begin{equation}
G_\pm(r) = r^{-1/4} F_\pm(r/\xi) [1 \pm  2^{-3/2} \xi^{-1}
] + {\cal O} (r^{-5/4})
\end{equation}
are correlation functions of the 2D Ising model, with scaling functions 
$F_\pm$ \cite{Wu76}. In Eq. (6) $\xi$ denotes the
spin--spin correlation length of the ladder system. 
The intra-- and inter--chain correlation functions differ only at short 
distances.
At large distances
$r/\xi \gg 1$, $C(r)$ decays as
\begin{eqnarray}
C(r) \sim r^{-\lambda} e^{- r/\xi}
\end{eqnarray} 
with $\lambda = 1/2$, which is equivalent to the 2D Ornstein--Zernike (OZ) form.
Figure 2 shows the low--$T$ correlation function for $n_c = 2$ at a weak 
coupling of $J_{\bot} = 0.2$.
The lines are the result of a fit to Eq. (5) with only two fitting parameters: 
$\xi$ and an overall amplitude.
The fit is excellent over the entire range, and we obtain $\xi = 19.4(2)$.
As is evident in Fig. 2, Eq. (5) correctly captures the crossover from the 
short-- to the long--distance behavior with a concomitant change in length 
scales by $\sim 3$.

\begin{figure}
\centerline{\epsfxsize=3.25in\epsfbox
{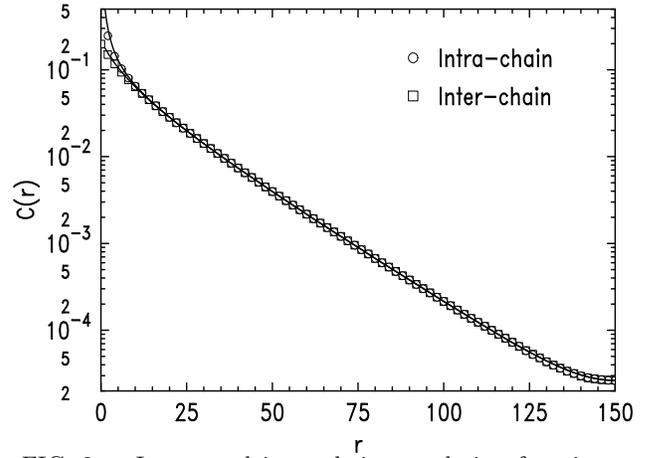}}
\caption{
Intra-- and inter--chain correlation functions at $T = 0.01$ for a $S=1/2$ 
two--chain Heisenberg ladder of length 300 with $J_{\bot} = 0.2$.
Note that periodic boundary conditions were employed along the chains and that 
for clarity $C(r)$ is shown only for even $r$.
The lines are the result of a fit to 
Eq. (5) in the symmetrized form 
$[C(r) + C(300 - r)].$
}
\end{figure}
In order to deduce $\xi(n_c,J_{\bot};T)$ for general $n_c, J_{\bot},$ and $T$, 
we fit $C(i,j)$ at large distances $r \geq 3 \xi$ to the asymptotic form 
Eq. (7).
For $n_c =4$ and 6, Eq. (7) with $\lambda = 0.5$ describes our low--$T$ data 
very well.
However, we find that the asymptotic behavior for even $n_c$ crosses over 
between
$T \approx 0.2 \Delta(n_c,J_{\bot})$ and $T \approx 0.4 \Delta(n_c,J_{\bot})$
to the 1D OZ form, that is, $\lambda = 0$.
For odd $n_c$ we find that the 1D form works very well at all 
temperatures.

In Fig. 3, the result of this analysis for $J_{\bot} = 1$ is shown together 
with $\xi(T)$ for the square lattice as obtained by both Monte 
Carlo \cite{THESIS} and neutron scattering in $Sr_2 Cu O_2 Cl_2$ \cite{SRCUOCL}.
Due to the presence of a gap for even $n_c$, $\xi(T)$ remains finite in the 
limit $T \rightarrow 0$.
We estimate that $\xi(n_c,1;0) = 3.24(5)$ and 10.3(1) for $n_c = 2$ 
and 4, respectively.
The result for $n_c =2$ is very close to the value $\xi = 3.19(1)$ obtained 
previously \cite{Whi94}.
However, in Ref.\cite{Whi94} $\xi = 5$ to 6 was obtained for $n_c =4$, which 
is significantly smaller than our result.
For $n_c = 2$ and 4 we obtain the respective velocities 
$c(n_c,1) = \Delta(n_c,1)\xi(n_c,1;0) = 1.63(2)$ and 1.65(3) which lie 
in between the 1D and 2D values $c_{1D} = \pi/2$ and $c_{2D} \simeq 1.68$.

The correlation length of the single $S=1/2$ chain has been determined in a 
thermal Bethe--ansatz study \cite{NY91}:
\begin{equation}
\xi^{-1}_{1D} (T) \simeq T \left[2 - b^{-1} \left( 1 - 0.486 b^{-1} \ln(b) 
\right) \right]
\end{equation}
with $b = -\ln (0.3733T)$.
For $T \leq 0.3$ our data for the chain ($n_c = 1$) agree with this low--$T$ 
form, indicated by the dashed line in Fig. 3.
We observe that with decreasing temperature $\xi$ for $n_c =3$ and 5 gradually 
approaches $\xi_{1D}$.
\begin{figure}
\centerline{\epsfxsize=3.25in\epsfbox
{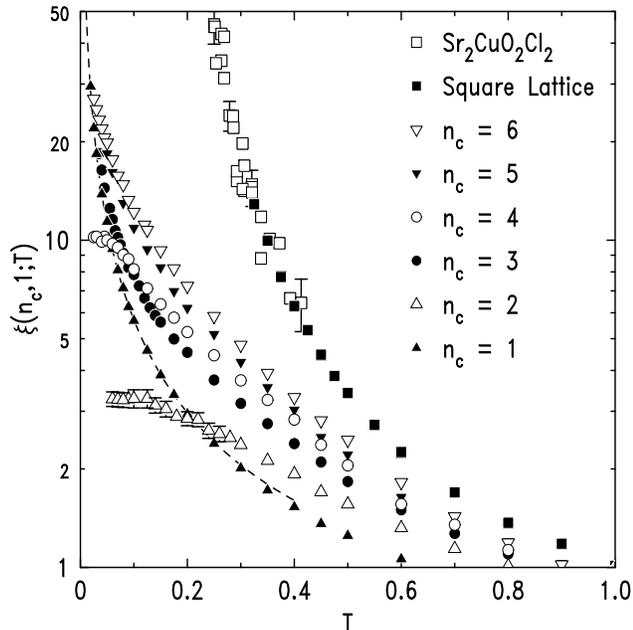}}
\caption{
Correlation length of $n_c$ isotropically coupled 
antiferromagnetic $S = 1/2$ chains. 
The result for the single chain ($n_c = 1$) is compared with the 
theoretical prediction Eq. (8) (dashed line).
Also shown are the results for the square lattice obtained from both Monte
Carlo simulations \protect\cite{THESIS} as well as neutron scattering 
experiments in $Sr_2CuO_2Cl_2$ \protect\cite{SRCUOCL}.
}
\end{figure}
 
In Fig. 4 the static structure factor at $(\pi,\pi)$,
$C_{\pi,\pi} = \sum_{i,j} C(i,j)$,
is shown for the isotropically coupled systems.
The overall trend with $n_c$ is similar to that for $\xi$.
However, $C_{\pi,\pi}(n_c,J_{\bot};T)$ for even $n_c$ exhibits a peak at a 
temperature $T_{max}$ well below the corresponding spin gap.
Not surprisingly, $T_{max}$ appears to coincide with the temperature at which 
the asymptotic correlation function Eq. (7) begins to cross over from the 
1D OZ form at higher temperatures to the low--$T$ 2D form.

Figure 5 shows the correlation length for the two--chain ladder at several 
inter--chain couplings $J_{\bot} \leq 1$.
For $T \approx \Delta(2,J_{\bot})$, $\xi(2,J_{\bot};T)$ is larger than 
$\xi_{1D} (T)$, since the ladder system has a higher effective 
coordination number than the chain.
Well below $T = \Delta(2,J_{\bot})$, the presence of the non--zero spin gap 
becomes apparent, and $\xi(2,J_{\bot};T \rightarrow 0)$ remains finite.
With decreasing $J_{\bot}$, the gap decreases as shown in Fig. 1, 
and correspondingly,  $\xi(2,J_{\bot};0)$ increases from 3.24(5) at 
$J_{\bot} = 1$ to 25.6(2) at $J_{\bot} = 0.15$.
In the weak--coupling regime, $J_{\bot} \leq 0.5$, the velocity 
$c(2,J_{\bot}) = \Delta(2,J_{\bot})\xi(2,J_{\bot};0)$ equals that of the 
single chain, $c_{1D} = \pi/2$, to within the error of our analysis.
In this regime the heuristic form 
\begin{equation}
\xi^{-1} (2,J_{\bot};T) = c_{1D}^{-1} \Delta(2,J_{\bot}) + \xi^{-1}_{1D} (T)
e^{-\Delta(2,J_{\bot})/T}
\end{equation}
describes our correlation length data very well, as shown by the solid lines 
in Fig. 5.
Note that since the spin gap has been determined from measurements of 
$\chi (T)$, this comparison contains no free parameters.
\begin{figure}
\centerline{\epsfxsize=3.25in\epsfbox
{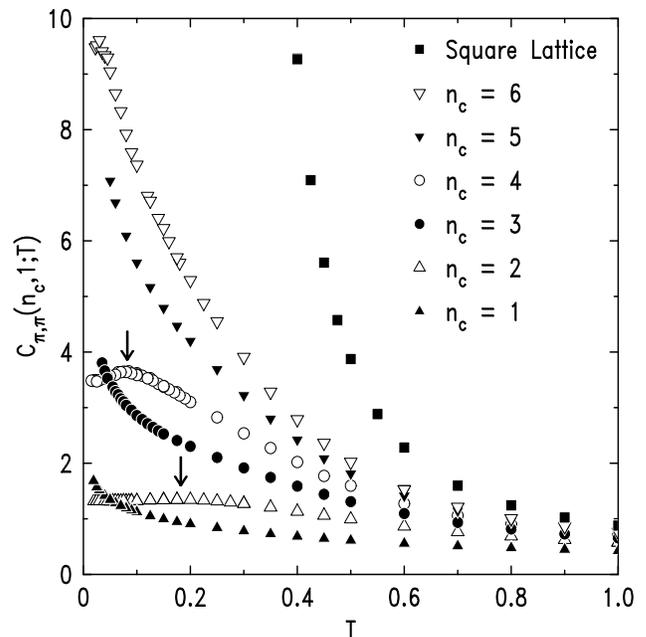}}
\caption{
Static structure factor at $(\pi,\pi)$ for isotropically 
coupled antiferromagnetic $S=1/2$ ladders, as well as for the single chain 
($n_c = 1$) and the square lattice \protect\cite{THESIS}.
For even $n_c$, $C_{\pi,\pi}$ exhibits a peak (indicated by arrows) at a 
temperature below that corresponding to the relevant spin gap.
}
\end{figure}

The temperature dependence of $\xi (2,J_{\bot};T)$ in the weak--coupling 
regime primarily results from that of the single chain, Eq. (8).
Apart from logarithmic corrections, the latter is simply 
$\xi_{1D} (T) \sim T^{-1}$.
Eq. (9) thus becomes 
$(\Delta \xi)^{-1} = 2/\pi + A (T/\Delta) e^{-\Delta/T}$,
which suggests plotting
$\Delta \xi$ versus $T/\Delta$ to test for the anticipated scaling for 
$J_{\bot} \leq 0.5$.
As shown in the inset of Fig. (5), our correlation length data indeed 
collapse onto a universal curve.
Over the indicated range the effective value for $A$ is $\sim 1.7$ compared 
to the low--$T$ value of 2.

For the two--chain ladder, we have established that 
$\Delta(2,J_{\bot} \rightarrow 0) = 0.41(1)J_{\bot}$.
We can furthermore estimate the weak--coupling behavior for $n_c = 4$.
From Fig. 1, it can be inferred that 
$\Delta(4,J_{\bot}) \simeq \Delta(2,J_{\bot}) - 0.35J_{\bot}$,
which leads to $\Delta(4,J_{\bot} \rightarrow 0) = 0.06J_{\bot}$ 
as a lower bound.
From the lowest $J_{\bot}$ for $n_c = 4$, we have 
$\Delta(4,0.6) = 0.12(1)J_{\bot}$ which constitutes an upper bound. 
Thus we arrive at the estimate 
$\Delta(4,J_{\bot} \rightarrow 0) = 0.09(3)J_{\bot}$.

We note that the trend with increasing $n_c$ of the ladder gap resembles that 
of the single integer--spin chain as a function of increasing $S$:
For $S=1$, $\Delta_{S = 1} = 0.4105J_{\|}$ is known very accurately \cite{S1}, 
while recent estimates for $S=2$ range between 
$\Delta_{S=2} = 0.049(18)J_{\|}$ and $0.085(5)J_{\|}$ \cite{S2}.
It has recently been argued that the two--chain ladder with 
antiferromagnetic inter--chain coupling lies in the same phase as the 
$S = 1$ single chain \cite{Whi96}, and that the weak--coupling physics is 
independent of the sign of $J_{\bot}$ \cite{She96}. 
Clearly, further numerical and analytical work to establish this 
correspondence is necessary.
\begin{figure}
\centerline{\epsfxsize=3.25in\epsfbox
{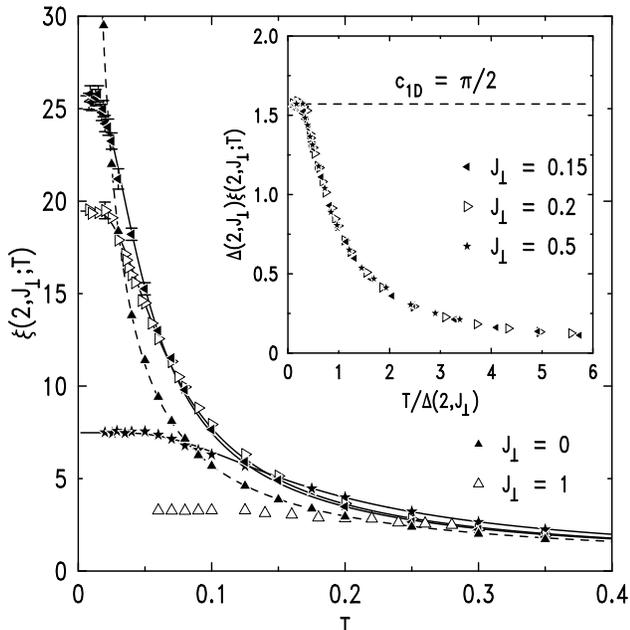}}
\caption{
Correlation length of two antiferromagnetically coupled $S = 1/2$ chains for
$J_{\bot} \leq 1$. 
The dashed line is Eq. (8), and
the solid lines are Eq. (9) with $\Delta (2,J_{\bot})$ as obtained from 
measurements of $\chi (T)$.
Inset: Scaling plot of $\Delta (2,J_{\bot})\xi(2,J_{\bot};T)$ versus 
$T/\Delta (2,J_{\bot})$ for $J_{\bot} \leq 0.5$.
}
\end{figure}
  
In summary, our Monte Carlo calculations on spin ladders have provided 
a number of new results.
First, our measured low--temperature correlation function for the 
weakly--coupled two--chain system agrees very well with the predicted 
theoretical form \cite{She96}, Eq. (5), which is claimed to be exact.
In particular, Eq. (5) captures the crossover in behavior between short 
and long distances.
With increasing temperature the even ladder correlations cross over from the 
2D Ornstein--Zernike form to the 1D form of pure exponential decay.
Second, for $J_{\bot} \leq 0.5$ we have discovered that 
$\Delta(2,J_{\bot})\xi(2,J_{\bot};T)$ 
exhibits a rather simple scaling behavior in the variable 
$T/\Delta(2,J_{\bot})$.
Third, the structure factor for even $n_c$ exhibits a peak at a temperature 
well below the corresponding spin gap.
Our results for the correlation length and the static structure factor of the 
isotropically coupled ladders pertain to the experimental systems. 
For the prototype $S=1/2$ square--lattice Heisenberg antiferromagnet 
$Sr_2CuO_2Cl_2$, neutron scattering measurements of the correlation length 
are in excellent agreement with Monte Carlo simulations and theory 
\cite{SRCUOCL}, as well as high--temperature series expansion \cite{Els95}.
Once sizable single crystals of the ladder systems become available, we 
intend to extend our neutron scattering work to these interesting systems.
We also hope that our Monte Carlo results will motivate future theoretical 
efforts to predict $C(i,j)$ for general $n_c, J_{\bot}$, and $T$.

We would like to thank R.J. Gooding, T.M. Rice, and D.J. Scalapino for 
valuable discussions, and the groups 
of P.A. Lee and X.--G. Wen at MIT for access to their computer.
This work was supported 
by the NSF under Grant No. DMR 93--15715, by the MRSEC Program
of the NSF under award number DMR 94--00334,
and by the DOE under cooperative research agreement 
No. DE--FC02--94ER40818.
U.--J.W. was also supported by an A.P. Sloan Fellowship.


\end{document}